\UseRawInputEncoding
\documentclass[pra,aps,twocolumn,showpacs]{revtex4}
\usepackage{times,epsfig,amssymb,amsfonts,amsmath}
\newcommand{\ket}[1]{|#1\rangle}
\newcommand{\bra}[1]{\langle #1|}
\newcommand{\proj}[1]{\ket{#1}\bra{#1}}
\usepackage{graphicx}
\usepackage{subfigure}
\usepackage{upgreek}

\begin{document}

\title{Long-range multipartite quantum correlations	and factorization in a one-dimensional spin-1/2 $XY$ chain}

\author{Lin-Lin Su$^1$}
\author{Jun Ren$^1$}
\email{renjun@hebtu.edu.cn}
\author{Z. D. Wang$^2$}
\email{zwang@hku.hk}
\author{Yan-Kui Bai$^{1,2}$}
\email{ykbai@semi.ac.cn}
\affiliation{$^1$ College of Physics and Hebei Key Laboratory of Photophysics Research and Application, Hebei Normal University, Shijiazhuang, Hebei 050024, People's Republic of China\\
$^2$ Guangdong-Hong Kong Joint Laboratory of Quantum Matter, Department of Physics, and HKU-UCAS Joint Institute for Theoretical and Computational Physics at Hong Kong, The University of Hong Kong, Pokfulam Road, Hong Kong, China}

\begin{abstract}
We study the properties of multipartite quantum correlation (MQC) in a one-dimensional spin-1/2 $XY$ chain, where the three-spin reduced states are focused on and the four introduced MQC measures are based on entanglement negativity and entanglement of formation. It is found that, even in the Ising case, the three-spin subsystems have the long-range MQCs and the tripartite quantum correlations beyond the nearest-neighbor three spins can detect the quantum phase transition and obey the finite-size scaling around the critical point. Furthermore, in the $XY$ model, we show that the two selected MQCs can indicate exactly the factorization point of the ground state for the anisotropic model in the thermodynamic and finite-size cases. Moreover, the spatial distribution of MQC based on entanglement negativity can attain to a much larger range by tuning the anisotropic parameter, and the newly defined MQC based on entanglement of formation can detect the bound entanglement in the three-spin subsystems when the entanglement negativity loses its efficacy.
\end{abstract}


\maketitle

\section{Introduction}

Quantum correlation \cite{rh09rmp,guh09prs,km12rmp,bru14rmp} is a kind of important physical resources and plays the crucial role in the tasks of quantum information processing, such as quantum secure communication and quantum computation. At the same time, quantum correlations also provide the effective tools for characterizing the properties of quantum many-body systems in the condensed matter physics \cite{ao02nat,tjo02pra,gv03prl,law04prl,sjg04prl,hl08prl,xc10prb,la08rmp,jens10rmp,aba19rmp}. For example, Osterloh \emph{et al} connected the critical phenomena \cite{sac00cam} with quantum entanglement for a class of magnetic systems \cite{ao02nat}, and showed that the nonanalyticity of energy can be manifested by the two-qubit concurrence \cite{woo98prl} in the nearest and the next-nearest spins at the critical point. In the last two decades, bipartite quantum correlations have been widely studied in quantum many-body systems, which helped develop the precise language for understanding quantum phase transition (QPT) in the interacting spin models.

Multipartite quantum correlation (MQC) can reveal more richer properties in the ground state of many-body systems (see a review paper \cite{gc18rpp} and references therein). In particular, it was shown that the MQC can exist and indicate the QPT even when the bipartite quantum correlations disappear \cite{tcw05pra,tro06pra,tro06prl,gia14njp,rrs21pra}.
Most of the existing studies are based on the MQC such as global entanglement \cite{mey02jmp}, geometric entanglement \cite{wei03pra}, residual entanglements \cite{vc00pra,osb06prl,yco07pra,yk14prl,yk09pra}, global quantum discord \cite{rul11pra}, multipartite nonlocality \cite{col02prl,ban11prl} and so on, where complete information about the ground state is needed to calculate the correlation functions in the interacting multipartite spin systems (for exemptions, see \cite{oru08prl,oru10prb,am10pra,bis14pra,zys15pra,bay17prl,fil17prb,rr18prb,sh20prb,jbx2021ol,
jbx2021pre,sun21pra,es22pra}). The MQCs have achieved great success in analyzing the QPTs for many-body systems, but the measurement of complete information on ground state is very difficult in general. Therefore, it is desirable to study the MQC in the reduced subsystems of the ground state, where on the one hand less information about the overall ground state is required and on the other hand more properties in the many-body system can be obtained in comparison to the two-site correlations. In this way, the main obstacle comes from the fact that the theory of MQC for multipartite mixed states is still not fully developed.

Based on the biseparable criterion $I_2$ \cite{hub10prl}, Giampaolo and Hiesmayr analyzed the relation between genuine tripartite entanglement and the QPT in the $XY$ model where the reduced state of three adjacent spins is considered \cite{smg13pra}. Similarly, utilizing genuine multipartite concurrence \cite{hie08pra,mzh11pra,has12pra}, the QPT and finite-size effects in the cluster-Ising model were studied via the reduced state of three central spins which has the specific $X$-form \cite{smg14njp}. According to genuine multipartite negativity \cite{guh11prl}, Hofmann \emph{et al} showed the existence of short-range multipartite entanglements in the reduced states of the $XY$ model and investigated the scaling property of tripartite entanglement in three nearest neighbor spins close to the QPT \cite{mh14prb}. However, knowledge about the MQC beyond the next-nearest case is still lacking, although the long-range multipartite entanglement close to the critical point in the $XXZ$ model was detected with the help of entanglement witness \cite{js14pra}. Moreover, it is still an open problem that whether or not the MQC beyond the adjacent three spins can detect the QPT and obey the finite-size scaling. In addition, a good candidate for the long-range MQCs should be effectively controlled and able to capture the interesting properties of many-body systems, for example, the factorization property of ground state in the anisotropic $XY$ model \cite{tr04prl,el61ap,pp70ap,bar70pra,be71pra}.

In this paper, we focus on the spatial distribution of MQCs in three-spin reduced states of a one-dimensional spin-1/2 $XY$ chain, and study the properties of criticality and factorization in the multipartite systems via the tripartite quantum correlations. The four intoduced MQC measures are based on entanglement negativity \cite{ver02pra} and the entanglement of formation \cite{ben96pra}, which are computable and have the larger spatial distributions beyond next-nearest-neighbor three spins. It is found that the tripartite quantum correlations in reduced subsystems can detect the ordered-nonordered transition of Ising model and obey the finite-size scaling even beyond the nearest neighbor cases. Furthermore, we show that the two selected MQCs can indicate exactly the factorization point of the anisotropic $XY$ chain in both the thermodynamic limit and the finite size case. In particular, it is revealed that the spatial distribution of MQC can be effectively modulated by the anisotropic parameter of the model, and the long-range bound entanglement \cite{wd00pra,dp07njp} can be discriminated via our newly defined MQC based on entanglement of formation.

This paper is organized as follows. In Sec.~II, we introduce the anisotropic $XY$ model and the four utilized MQCs. Next, the spatial distribution of MQCs and their critical behaviors in the Ising case are studied in Sec.~III. In Sec.~IV, the MQC modulation, the factorization property, and bound entanglement in the $XY$ chain are investigated. Finally, some discussions and a brief summary of main results are given in Sec.~V.

\section{The model and multipartite quantum correlations}

\subsection{1D $XY$ model and the analytical form of 3-qubit reduced state for the ground state}

We now consider a spin-1/2 $XY$ chain under the transverse magnetic field on $L$ spins, where the interactions involve only nearest-neighbor couplings and the Hamiltonian with periodic boundary conditions is given by \cite{bar70pra,mh14prb}
\begin{equation}\label{1}
H=-\lambda\underset{i=1}{\overset{L}{\sum}}[\frac{1+\gamma}{2}\sigma^{x}_{i}\sigma^{x}_{i+1}+\frac{1-\gamma}{2}\sigma^{y}_{i}\sigma^{y}_{i+1}]
+\underset{i=1}{\overset{L}{\sum}}\sigma^{z}_{i},
\end{equation}
where the $\lambda\geq0$ is the interaction strength with respect to the external magnetic field, and $\{\sigma^{x}_{i}$, $\sigma^{y}_{i}$, $\sigma^{z}_{i}\}$ are the Pauli operators on the $i$th spin site. The parameter $\gamma$ ranging from 0 to 1 represents the anisotropic property of the system, and the chain is referred to as the Ising model for $\gamma=1$ and isotropic $XY$ model for $\gamma=0$. In the thermodynamic limit ($L\rightarrow \infty$), the ground state of the many-body systems undergoes a quantum phase transition at the critical point $\lambda_{\rm c}=1$. At the phase transition, the correlation length diverges as $\xi \sim 1/|\lambda-\lambda_{\rm c}|$ \cite{bar70pra,be71pra}, but it was pointed out that the two-spin entanglement length is short-ranged \cite{ao02nat}. In this work, we will focus on the spatial distribution of MQCs and the related critical phenomena in the spin model.

The $XY$ chain is one of the few models for which the ground state and its reduced states can be solved analytically with an arbitrary chain length. Here, we consider 3-qubit reduced density matrix $\rho_{ijk}$ of the ground state, in which the subscripts denote the $i$th, $j$th and $k$th sites in the spin chain respectively. Due to the translation invariance property of the system, the reduced state can be rewritten as $\rho_{i-\alpha,i,i+\beta}$ and the spin arrangement index $m=(\alpha,\beta)$ uniquely determine the form of the reduced density matrix no matter what the value of the index $i$ is chosen to be. Moreover, according to the mirror symmetry, the indices $(\alpha,\beta)$ and $(\beta,\alpha)$ lead to the same form of the tripartite reduced states. Using the method introduced in Ref.~\cite{el61ap}, the Hamiltonian in Eq.~(1) can be exactly diagonalized and the 3-qubit reduced state of the ground state $\ket{\psi}$ is expressed as \cite{dp07njp}
\begin{eqnarray}\label{2}
\rho_{ijk}(\alpha, \beta)&=&\rho_{i-\alpha,i,i+\beta}\nonumber\\
 &=&\frac{1}{8}{\underset{p,q,s}{\sum}}\langle\sigma^{p}_{i-\alpha}\sigma^{q}_{i}\sigma^{s}_{i+\beta}\rangle_{\ket{\psi}}
 \sigma^{p}_{i-\alpha}\sigma^{q}_{i}\sigma^{s}_{i+\beta},
\end{eqnarray}
where each of the summations $p,q,s$ runs over $\{x,y,z,0\}$, and the expectations of Pauli operators act on the ground state $\ket{\psi}$ of the whole chain. A brief review of the diagonalization process and the detailed expressions for the three-qubit reduced state in the finite and infinite chains are presented in Appendix A.

\subsection{Multipartite quantum correlations based on entanglement negativity}

Negativity is a computable entanglement measure for bipartite mixed states $\rho_{AB}$, which is defined as $N(\rho_{AB})=||\rho^{T_A}_{AB}||_1-1=\sum_k|\lambda_k|-1$ with $\lambda_ks$ being the eigenvalues of partial transposition matrix $\rho^{T_A}_{AB}$ \cite{ver02pra}. In addition, the logarithmic negativity $LN(\rho_{AB})=\mbox{Log}_2||\rho_{AB}^{T_A}||_1$ is the upper bound of the distillable entanglement \cite{ple05prl}. Bennett \emph{et al} proposed three reasonable postulates for measures of genuine multipartite correlation and pointed out that a state of $N$ particles has genuine $n$-partite correlations if it is nonproduct in every bipartite cut \cite{ben11pra}. Therefore, via the bipartite entanglement negativity, we can define a measure for tripartite quantum correlation
\begin{equation}\label{3}
	N_3(\rho_{ijk})=[N(\rho_{i|jk}) \cdot N(\rho_{j|ik}) \cdot N(\rho_{k|ij})]^{1/3},
\end{equation}
where $\rho_{ijk}$ is a tripartite mixed state, and the bipartite negativities characterize the quantum correlations in different bipartite partitions. According to the definition, we have the MQC $N_3(\rho_{ijk})$ being zero when the tripartite mixed state is biseparable in any partition. It is noted that a similar measure for multipartite pure state was utilized to characterize genuine tripartite entanglement \cite{bay17prl,sc10pra}.

Entanglement monogamy is one of the most important properties in many-body systems \cite{rh09rmp}. Coffman, Kundu, and Wootters proved the first quantitative relation for the squared concurrence in three-qubit systems and showed that the residual entanglement is a genuine tripartite entanglement measure \cite{vc00pra}. Ou and Fan further proved that entanglement negativity is monogamous in $N$-qubit pure states \cite{yco07pra}, and the case $N^2_{A|BC}\geq N^2_{AB}+N^2_{AC}$ in high-dimensional tripartite systems was also verified numerically \cite{hh15pra,ba16prb,bay17prl}. Moreover, it has been shown that, utilizing the residual entanglement of squared negativity in mixed states of many-body systems, one can detect the MQC and distinguish between the frustrated and nonfrustrated regimes in spin chains \cite{bai06pra,rao13pra,ren21arx}. Based on the monogamy of negativity, a genuine tripartite quantum correlation in a three-qubit mixed state $\rho_{ijk}$ can be defined as
\begin{equation}
	T_3(\rho_{ijk})={\rm max}\left\lbrace \left( T^{i}_3+T^{j}_3+T^{k}_3\right) /3,0\right\rbrace,
\end{equation}
where $	T^{i}_3=N^2(\rho_{i|jk})-N^2(\rho_{ij})-N^2(\rho_{ik})$ is the residual entanglement corresponding to the central qubit $i$ and the meanings for $T^{j}_3$ and $T^{k}_3$ are similar. The nonzero $T_3$ characterizes the tripartite quantum correlation which cannot be restored in two-qubit subsystems.

\subsection{Multipartite quantum correlations based on entanglement of formation}

The entanglement of formation is a well-defined bipartite entanglement measure, which has the operational meaning in quantum state preparation and data storage \cite{ben96pra}. It was proved that the squared entanglement of formation is monogamous in multiqubit systems, and its residual entanglements can characterize genuine multipartite entanglement \cite{yk14prl,bai13pra,bai14pra}. In particular, for a three-qubit mixed state $\rho_{ijk}$, the residual entanglement can indicate genuine tripartite quantum correlation, which can be expressed as \cite{yk14prl}
\begin{equation}\label{5}
\tau^{i}_{\rm SEF}(\rho_{ijk})=E^2_f(\rho_{i|jk})-E^2_f(\rho_{ij})-E^2_f(\rho_{ik}),
\end{equation}
where $\tau^{i}_{\rm SEF}(\rho_{ijk})$ is the tripartite quantum correlation corresponding to the central qubit $i$, and $E_f(\rho_{ir})$ is the bipartite entanglement between the qubit $i$ and the subsystem $r$ with $r\in\{j, k, jk\}$. The entanglement of formation between qubit $i$ and qubit pair $jk$ is defined as $E_f(\rho_{i|jk})=\mbox{min}\sum_{s}p_s E_f(\ket{\psi^s}_{i|jk})$ in which $E_f(\ket{\psi^s}_{i|jk})=S(\rho_i^s)=-\mbox{Tr}\rho_i^s\mbox{log}_2\rho_i^s$ is the von Neumann entropy and the minimum runs over all the pure state decompositions $\rho_{ijk}=\sum_s p_s \proj{\psi^s_{ijk}}$ \cite{ben96pra}. An analytical formula for two qubit entanglement of formation was given by Wootters \cite{woo98prl} $E_f(\rho_{ij})=h[(1+(1-C_{ij}^2)^{1/2})/2]$, in which  $h(x)=-x\mbox{log}_2x-(1-x)\mbox{log}_2(1-x)$ is the binary entropy and $C_{ij}=\mbox{max}[0,\sqrt{\lambda_1}-\sqrt{\lambda_2}-\sqrt{\lambda_3}-\sqrt{\lambda_4}]$ is the concurrence with $\lambda_i$ being the decreasing eigenvalues of matrix $\rho_{ij}(\sigma_y\otimes\sigma_y)\rho_{ij}^*(\sigma_y\otimes\sigma_y)$. The $\tau^{i}_{\rm SEF}(\rho_{ijk})$ is an effective MQC indicator which can characterize genuine tripartite quantum correlation dynamics in multipartite cavity-reservoir systems \cite{yk14prl}, and the cases for MQCs $\tau^{j}_{\rm SEF}(\rho_{ijk})$ and $\tau^{k}_{\rm SEF}(\rho_{ijk})$ are similar. However, although the residual entanglement of formation has strong detection ability for the MQC, its computation for a generic mixed state is very difficult except for some specific mixed states such as rank-2 mixed states. Therefore, the computable upper and lower bounds for the MQC based on the residual entanglement of formation are desirable.

Next, we first introduce a computable upper bound for the tripartite quantum correlation $\tau^{i}_{\rm SEF}(\rho_{ijk})$ in a generic three-qubit mixed state. Recently, Wang and Wilde determined the PPT (positive partial transposition) exact entanglement cost \cite{aud03prl} of an arbitrary bipartite mixed state via the $\kappa$ entanglement \cite{xw20prl}. For any three-qubit mixed state $\rho_{ijk}$, the bipartite PPT exact entanglement cost between qubit $i$ and qubit pair $jk$ can be written as \cite{xw20prl}
\begin{eqnarray}\label{6}
&&E_{\mbox{PPT}}(\rho_{i|jk})=E_\kappa(\rho_{i|jk})\nonumber\\
&=&\log_2\underset{S_{i|jk}\geq0}{\inf}\lbrace {\rm Tr} [S_{i|jk}]: -S^{T_i}_{i|jk}\leq\rho^{T_i}_{i|jk}\leq S^{T_i}_{i|jk}\rbrace,
\end{eqnarray}
where the $\kappa$ entanglement $E_{\kappa}$ can be computed by means of a semidefinite program (SDP) \cite{lv96sr,github}. Furthermore, for the bipartite entanglement cost, we have the relation
\begin{equation}\label{7}
E_{\mbox{PPT}}(\rho_{i|jk})\geq LN(\rho_{i|jk})\geq E_{\alpha}(\rho_{i|jk})\geq E_f(\rho_{i|jk}),
\end{equation}
where the first inequality is due to the logarithmic negativity \cite{ple05prl} is the lower bound of the PPT entanglement cost \cite{aud03prl,xw20pra}, the second inequality comes from the relation between logarithmic negativity and R\'{e}nyi-$\alpha$ entropy entanglement for the case $1/2\leq \alpha<1$ \cite{song16sr,kim10jpa}, and in the last inequality we use the monotonic property of $E_\alpha$ along with the parameter $\alpha$ and the entanglement degenerating to the $E_f$ when $\alpha$ tends to 1. According to the established relation in Eq. (7), we obtain that the PPT exact entanglement cost is the upper bound for the entanglement of formation. Therefore, combining Eq. (5) and Eq. (7), we can define a computable upper bound of genuine tripartite quantum correlation
\begin{equation}\label{8}
\tau^{\rm UB}_{\rm SEF}(\rho_{ijk})=
\left( \tau^{i(\rm UB)}_{\rm SEF}+\tau^{j(\rm UB)}_{\rm SEF}+\tau^{k(\rm UB)}_{\rm SEF}\right) /3,
\end{equation}
where $\tau^{i(\rm UB)}_{\rm SEF}=E^2_{\rm PPT}(\rho_{i|jk})- E^2_f(\rho_{ij})- E^2_f(\rho_{ik})$ and the meanings of $\tau^{j(\rm UB)}_{\rm SEF}$ and $\tau^{k(\rm UB)}_{\rm SEF}$ with different central qubits are similar.

On the other hand, about the lower bound of genuine tripartite quantum correlation, the key step is the characterization on the lower bound of bipartite entanglement of formation in a three-qubit mixed state. Chen \emph{et al} proved an analytical lower bound for the entanglement of formation in arbitrary bipartite mixed state \cite{kc05prl}, and, for a three-qubit state $\rho_{ijk}$, the formula can be expressed as
\begin{eqnarray}\label{9}
E_f^{\rm LB}(\rho_{i|jk})=
\begin{cases}
0, \quad \Lambda=1,\\
H_2[(1+\sqrt{\Gamma})/2],\quad \Lambda\in[1,2],
\end{cases}
\end{eqnarray}
where $\Lambda=\max(||\rho^{T_i}_{ijk}||_1,||R(\rho_{ijk})||_1)$ is the maximum of trace norms between partial transposition matrix and realignment matrix, and $H_2(\cdot)$ is the binary entropy function with $\Gamma=1-(\Lambda-1)^2$. With the help of the lower bound of entanglement of formation, we can obtain an effective and computable lower bound of genuine tripartite quantum correlation
\begin{equation}\label{10}
\tau^{\rm LB}_{\rm SEF}(\rho_{ijk})=
{\rm max}\left\lbrace \left( \tau^{i(\rm LB)}_{\rm SEF}+\tau^{j(\rm LB)}_{\rm SEF}+\tau^{k(\rm LB)}_{\rm SEF}\right) /3,0\right\rbrace,
\end{equation}
where $\tau^{i(\rm LB)}_{\rm SEF}=[E^{\rm LB}_{f}(\rho_{i|jk})]^2- E^2_f(\rho_{ij})- E^2_f(\rho_{ik})$ characterizes the MQC corresponding to central qubit $i$, and tripartite quantum correlations $\tau^{j(\rm LB)}_{\rm SEF}$ and $\tau^{k(\rm LB)}_{\rm SEF}$ with different central qubits have the similar meanings.

\section{spatial distribution of MQCs and quantum phase transition in the Ising chain}

We first analyze the spatial distribution of MQCs in the spin-1/2 chain as shown in Eq. (1) and consider the case of $\gamma=1$, which corresponds to the transverse field Ising model. According to the expression of $\rho_{ijk}$ in Eq. (2), we know that an arbitrary three-spin reduced state of the ground state can be labeled by the index $m=(\alpha,\beta)$ which represents the spatial arrangement of the three spins. It was pointed out in the $XY$ model at the thermodynamic limit ($L\rightarrow\infty$) that the three-spin entanglements, characterized by permutation operators, multipartite concurrence and genuine negativity, can attain to the spatial distributions $m=(1,1)$ and $m=(2,1)$ \cite{smg13pra,smg14njp,mh14prb}. In regard to the four tripartite quantum correlations introduced in Sec. II, it is desirable to investigate that whether the spatial distributions of the MQCs have the longer ranges.

\begin{figure}
	\epsfig{figure=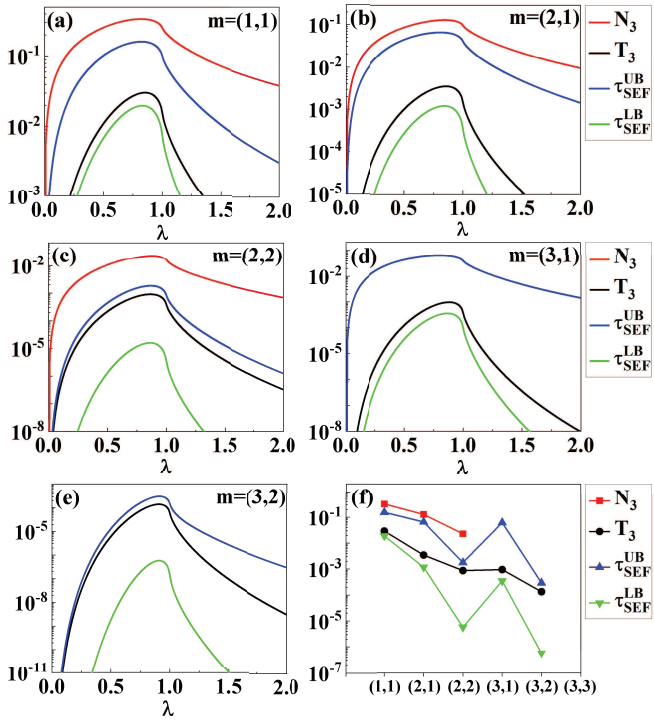,width=0.48\textwidth}
\caption{(Color online) Spatial distributions of four kinds of tripartite quantum correlations in three-spin reduced states of the ground state for the Ising model at the thermodynamic limit. (a)-(e) MQCs along with the parameter $\lambda$ for different spatial distributions $m=(1,1)$, $(2,1)$, $(2,2)$, $(3,1)$ and $(3,2)$, respectively. (f) The maxima of four kinds of MQCs in different distributions, and all the tripartite quantum correlations disappear in the length $m=(3,3)$.}
\end{figure}

We first calculate the reduced state $\rho_{ijk}$ of the ground state for the Ising model with the limit $L\rightarrow\infty$, and then evaluate the four kinds of tripartite quantum correlations $N_3(\rho_{ijk})$, $T_3(\rho_{ijk})$, $\tau_{\rm SEF}^{\rm UB}(\rho_{ijk})$, and $\tau_{\rm SEF}^{\rm LB}(\rho_{ijk})$, respectively. As shown in Fig. 1, we plot the four MQCs of three spins with the spatial distributions $m=(1,1)$, $(2,1)$, $(2,2)$, $(3,1)$ and $(3,2)$. In Figs. 1(a)-1(c), the tripartite quantum correlation $N_3(\rho_{ijk})$ (the red-solid line) has the maximal values, but the MQC vanishes away when the distribution of three spins goes beyond $m=(2,2)$. For the distributions $m=(3,1)$ and $m=(3,2)$ in Figs. 1(d) and 1(e), the tripartite quantum correlation $\tau_{\rm SEF}^{\rm UB}(\rho_{ijk})$ (the blue-solid line) is maximal. Moreover, in all the distributions, the MQCs $T_3(\rho_{ijk})$ (the black-solid line) and $\tau_{\rm SEF}^{\rm LB}(\rho_{ijk})$ (the green-solid line) are nonzero but have the smaller values. In Fig.  1(f), we plot the maximums of the four MQCs in different spatial distributions where the tripartite quantum correlation $N_3$ has the maximal value but the smallest correlation length, and all the MQCs disappear at the spatial distribution $m=(3,3)$ in this Ising system. It is clear that the spatial distributions of these MQCs have the larger correlation length than those of tripartite quantum entanglement \cite{smg13pra,smg14njp,mh14prb} not exceeding $m=(2,1)$.

\begin{figure}
	\epsfig{figure=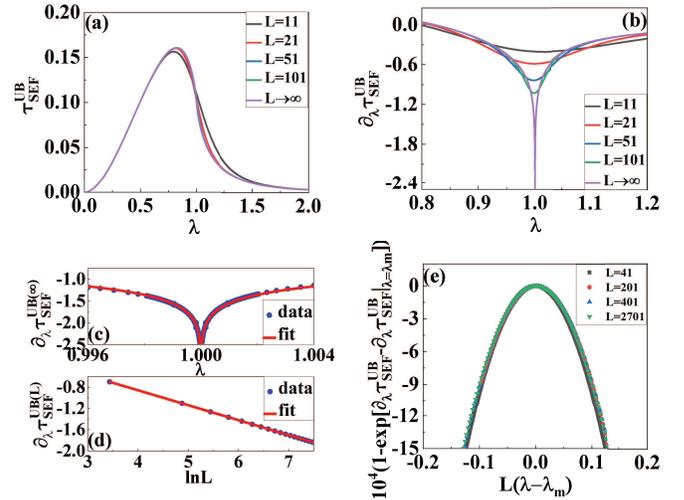,width=0.48\textwidth}
	\caption{(Color online) The critical phenomenon and finite-size effects in the Ising system characterized by the tripartite quantum correlation $\tau^{\rm UB}_{\rm SEF}(\rho_{ijk})$ with the spatial distribution $m=(1,1)$. (a) The change of the MQC along with the parameter $\lambda$ for different chain lengths. (b) The first-order derivative $\partial_{\lambda}\tau^{\rm UB}_{\rm SEF}$ versus the parameter $\lambda$ for different chain sizes. (c) The minimum of derivative $\partial_{\lambda}\tau^{\rm UB(\infty)}_{\rm SEF}$ versus the parameter $\lambda$ close to the critical point $\lambda_c=1$. (d) The finite-size logarithmical scaling of the derivative $\partial_{\lambda}\tau^{\rm UB(L)}_{\rm SEF}$ versus the chain length. (e) The universality of the tripartite quantum correlation $\tau^{\rm UB}_{\rm SEF}$ and the homogeneous function for different chain sizes.}
\end{figure}

Next, based on the computable tripartite quantum correlations, we analyze the critical phenomenon and finite-size effects in the Ising system. In Fig. 2, we characterize these properties by the correlation $\tau^{\rm UB}_{\rm SEF}(\rho_{ijk})$ in three adjacent spins with the spatial distribution $m=(1,1)$. The change of tripartite quantum correlation along with the parameter $\lambda$ is plotted for different chain lengths as shown in Fig. 2(a), where the curve for the short chain length $L=11$ (the black-solid line) has a little deviation from the case $L\rightarrow \infty$ (the purple-solid line). It is well known that the $XY$ spin model given in Eq.~(1) will undergo a quantum phase transition at the critical point $\lambda_{\rm c}=1$, and the previous study showed that the nearest-neighbor three-spin entanglement based on genuine multipartite negativity can indicate this transition and the related finite-size effects \cite{mh14prb}. In Fig.2 (b), the first-order derivative $\partial_{\lambda}\tau^{\rm UB}_{\rm SEF}(\rho_{ijk})$ with different chain lengths is plotted, which indicates the QPT and diverges at the critical point $\lambda_c=1$ for the thermodynamic case $L\rightarrow \infty$. Moreover, as shown in the figure, there is also distinct minimum of $\partial_{\lambda}\tau^{\rm UB(L)}_{\rm SEF}$ for finite system size $L$ at the pseudo-critical point $\lambda_{\rm m}(L)$, which approaches the critical point like $\lambda_{\rm m}(L)-\lambda_{\rm c}\sim -L^{-1.38}$. At the thermodynamic limit, the derivative $\partial_{\lambda}\tau^{\rm UB(\infty)}_{\rm SEF}$ around the critical point can be written as a function of $(\lambda-\lambda_{\rm c})$, and after fitting the expected behavior to our data we can obtain
\begin{equation}\label{11}
	\partial_{\lambda}\tau^{\rm UB(\infty)}_{\rm SEF}=0.2819 \ln|\lambda-\lambda_{\rm c}|+0.3906,
\end{equation}
which is depicted in Fig. 2(c). For sufficiently large $L$,
we also have a finite-size scaling relation
\begin{equation}\label{12}
	\partial_{\lambda}\tau^{\rm UB(L)}_{\rm SEF}[\lambda_{\rm m}(L)]=-0.2819 \ln L+0.2720,
\end{equation}
as shown in Fig. 2(d). Furthermore, we check the universality of the tripartite quantum correlation $\tau^{\rm UB}_{\rm SEF}$ by plotting the finite-size scaling. In the critical regime, we take a proper scaling \cite{mn1983} and analyze the distance of the minimum of $\tau^{\rm UB}_{\rm SEF}$ from the pseudo-critical point $\lambda_{\rm m}(L)$. A general relation can be obtained
\begin{equation}\label{13}
	\partial_{\lambda}\tau^{\rm UB}_{\rm SEF}-\partial_{\lambda}\tau^{\rm UB}_{\rm SEF}|_{\lambda=\lambda_{\rm m}}=\mathcal{R}_{\tau^{\rm UB}_{\rm SEF}}[L(\lambda-\lambda_{\rm m})],
\end{equation}
where $\mathcal{R}_{\tau^{\rm UB}_{\rm SEF}}(\cdot)$ is a homogeneous function for the MQC. As shown in Fig. 2(e), we plot the homogeneous curve via the data of chain lengths including $L=41,201,401$ and $2701$. In addition, we also performed the same analysis on the correlation $\tau^{\rm UB}_{\rm SEF}(\rho_{ijk})$ with the spatial distributions $m=(2,1), (2,2), (3,1)$, and observe the similar qualitative results in the Ising system. In Appendix B, the fitted relations for the case $m=(3,1)$ are given.

The tripartite quantum correlation $N_3(\rho_{ijk})$ has the relatively larger values than those of the other MQCs even beyond the nearest-neighbor case as show in Fig. 1(f). Here, we further analyze the properties of the Ising spin chain by utilizing the correlation $N_3(\rho_{ijk})$ with the spatial distribution $m=(2,1)$. In Fig. 3(a), the correlation  $N_3(\rho_{ijk})$ versus the parameter $\lambda$ is plotted in which the cases of short chains have a few deviation from that of the thermodynamic limit $L\rightarrow \infty$. In Fig. 3(b), the derivative $\partial_{\lambda}N^{(\infty)}_3$ can indicate the QPT at the point $\lambda_{\rm c}=1$, and the one for finite sizes $\partial_{\lambda}N^{(L)}_3$ has the minimum at the pseudo-critical point $\lambda_{\rm m}(L)$. Similarly, after fitting the data, we can obtain the relations about the derivative $\partial_{\lambda}N^{(\infty)}_3$ close to $\lambda_{\rm c}$ and finite-size scaling of $\partial_{\lambda}N^{(L)}_3$
\begin{eqnarray}\label{14}
&&\partial_{\lambda}N^{(\infty)}_3=0.1961 \ln|\lambda-\lambda_{\rm c}|+0.3042,\\
&&\partial_{\lambda}N^{(L)}_3[\lambda_{\rm m}(L)]=-0.1961 \ln L+0.2198,
\end{eqnarray}
where $\lambda_{\rm m}(L)-\lambda_{\rm c}\sim L^{-1.89}$ and the two equations are plotted in Figs. 3(c) and 3(d), respectively. In the quantum critical regime, the scaling behavior can be represented by
$\partial_{\lambda}N_3-\partial_{\lambda}N_3|_{\lambda=\lambda_{\rm m}}=\mathcal{R}_{N_3}[L(\lambda-\lambda_{\rm m})]$, for which the homogeneous function is plotted in Fig. 3(e) via the data of spin chain lengths $L=41,201,401$ and $2701$. About the correlation $N_3$ with the distributions $m=(1,1)$ and $m=(2,2)$, we can obtain the similar qualitative results in the Ising model. Moreover, the tripartite correlations $T_3(\rho_{ijk})$ and $\tau_{\rm SEF}^{\rm (LB)}(\rho_{ijk})$ can also characterize the critical phenomenon and finite-size scaling properties, and the details for their behaviors with the spatial distribution $m=(1,1)$ are presented in Appendix B.

\begin{figure}
	\epsfig{figure=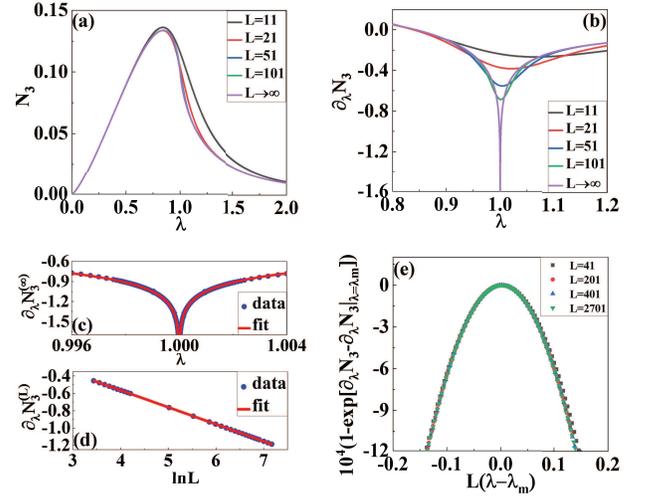,width=0.45\textwidth}
	\caption{(Color online) The critical phenomenon and finite-size scaling properties characterized by the tripartite quantum correlation $N_3(\rho_{ijk})$ with the spatial distribution $m=(2,1)$, including (a) the MQC with different chain lengths, (b) the derivative $\partial_{\lambda}N_3$ versus the parameter $\lambda$, (c) the fitting relation between $\partial_{\lambda}N_3^{(\infty)}$ and the parameter $\lambda$ close to the critical point, (d) the finite-size logarithmical scaling of $\partial_{\lambda}N_3^{(\rm L)}$, and (e) the homogeneous function for different chain lengths.}
\end{figure}

In this section, we have studied the spatial distribution of the four kinds of MQCs and their critical behaviors in the Ising system. In comparison with the previous study of tripartite quantum entanglement, the distribution of the MQCs can be extended to the longer range $m=(3,2)$. Furthermore, the four utilized tripartite quantum correlations, even beyond the adjacent three-spin case, can effectively characterize the critical phenomenon and finite-size scaling in the Ising system. In next section, we will further analyze the properties of the MQCs in the anisotropic $XY$ model.

\section{The modulation of MQC distribution and the factorization property in the XY chain}

We now turn to study the MQCs in the $XY$ model, for which the Hamiltonian as shown in Eq. (1) with the parameter $\gamma\neq 1$ being the the anisotropic parameter. Factorization is one of the most important properties in the anisotropic $XY$ model, which can be described as that the quantum correlation properties of the spin system undergo a sudden change in the ordered phase and the ground state in the thermodynamic limit turns into a fully factorized state at the factorization point \cite{smg08prl,smg09prb,sc13pra}
\begin{equation}\label{16}
\lambda_f=\frac{1}{\sqrt{1-\gamma^2}}.
\end{equation}
In the last section, we have found that all the four kinds of tripartite quantum correlations $N_3$, $\tau_{\rm SEF}^{\rm UB}$, $T_3$, and $\tau_{\rm SEF}^{\rm LB}$ can capture the properties of the QPT. But a desirable MQC measure should be able to capture more properties of the many-body systems, such as the factorization property in the $XY$ model.

\begin{figure}
	\epsfig{figure=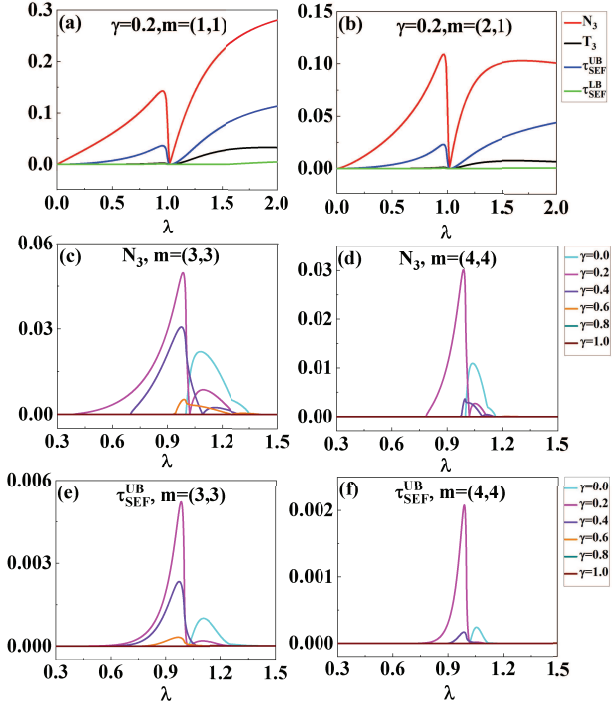,width=0.45\textwidth}
	\caption{(Color online) The MQCs versus the parameter $\lambda$ in the $XY$ model with $L\rightarrow \infty$. (a) and (b): Four kinds of MQCs with the anisotropic parameter $\gamma=0.2$ for the spatial distributions $m=(1,1)$ and $m=(2,1)$, respectively. (c) and (d): The correlation $N_3$ with different values of $\gamma$ for the distributions $m=(3,3)$ and $m=(4,4)$. (e) and (f): The correlation $\tau^{\rm UB}_{\rm SEF}$ with different values of $\gamma$ for the distributions $m=(3,3)$ and $m=(4,4)$, respectively.}
\end{figure}

For the $XY$ model with the anisotropic parameter $\gamma=0.2$ in the thermodynamic limit, we plot the four kinds of MQCs in the three-spin reduced state $\rho_{ijk}$ of the ground state as shown in Fig. 4(a) and Fig. 4(b), where the spatial distribution of the spins are chosen to be $m=(1,1)$ and $m=(2,1)$, respectively. In Fig. 4(a), we find that the correlations $N_3$ (the red-solid line) and $\tau^{\rm UB}_{\rm SEF}$ (the blue-solid line) can capture the sudden change of the factorization and pinpoint correctly the factorization point $\lambda_f=1/\sqrt{1-\gamma^2}\simeq 1.0206$, but the correlations $T_3$ (the black-solid line) and $\tau^{\rm LB}_{\rm SEF}$ (the green-solid line) cannot indicate the sudden change property of the ground state. In Fig. 4(b), the identification abilities for $N_3$ and $\tau^{\rm UB}_{\rm SEF}$ with the distribution $m=(2,1)$ are similar, but the other two MQCs still do not possess this property. Moreover, after comparing the evolutions of tripartite quantum correlations $N_3$ and $\tau^{\rm UB}_{\rm SEF}$ with those of the Ising cases shown in Fig. 2(a) and Fig. 3(a), we find that the correlation evolutions in the $XY$ model is quite different. Besides the factorization property, the anisotropic parameter $\gamma=0.2$ increases $N_3$ and $\tau^{\rm UB}_{\rm SEF}$ obviously in the ordered phase ($\lambda>1$) which means that the parameter $\gamma$ can modulate effectively the MQCs.

\begin{figure}
	\epsfig{figure=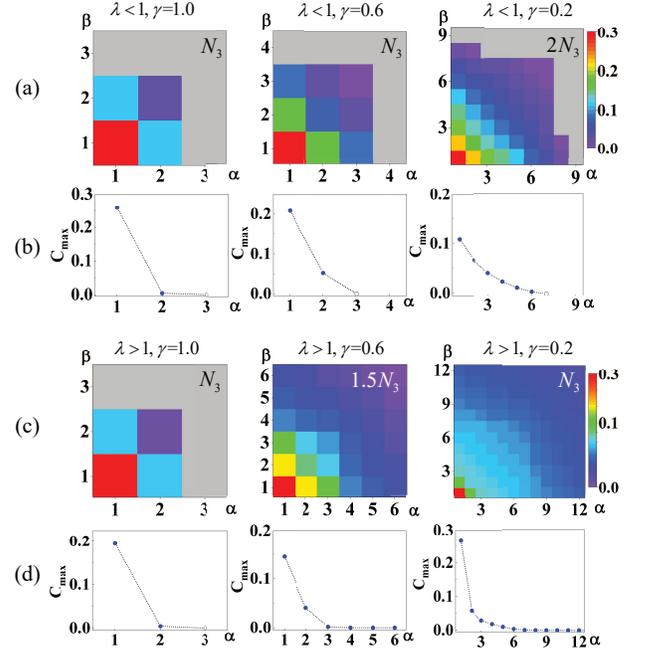, width=0.45\textwidth}
	\caption{(Color online) The effective modulation of spatial distributions of $N_3(\alpha, \beta)$ and $C_{\rm max}(\alpha)$ in the nonordered and ordered phases via the parameter $\gamma$ in the $XY$ model. (a) The distribution $N_3(\alpha, \beta)$ with $\gamma=1.0, 0.6$ and $0.2$ in the nonordered phase ($\lambda<1$). (b) The distribution $C_{\rm max}(\alpha)$ with the same values of $\gamma$ in the nonordered phase. The similar distributions but in the ordered phase ($\lambda>1$) are plotted in figures (c) and (d), respectively.}
\end{figure}

In the Ising system studied in the last section, the maximum of spatial distribution for the four kinds of MQCs is $m=(3,2)$, and the correlation $N_3$ disappears even beyond $m=(2,2)$ as shown in Fig. 1(f). Next, for the $XY$ system, it is natural to investigate the modulation effect of the parameter $\gamma$ on the spatial distribution of the tripartite quantum correlations $N_3$ and $\tau^{\rm UB}_{\rm SEF}$. We plot the correlation $N_3$ versus the parameter $\lambda$ with the spatial distributions $m=(3,3)$, $m=(4,4)$ in Fig. 4(c) and Fig. 4(d) respectively, and the anisotropic parameter $\gamma$ is chosen to be $0, 0.2, 0.4, 0.6, 0.8$, and $1.0$. It can be found that the tripartite quantum correlation $N_3$ is nonzero for most values of $\gamma$ in comparison with the zero $N_3$ for the Ising case $\gamma=1$, which illustrates that the anisotropic parameter $\gamma$ can modulate the MQC to a longer range of spatial distribution. Moreover, compared the $N_3$ in Fig. 4(a)-(d), we find that the MQC in the ordered phase ($\lambda>1$) is sensitive to the spatial distribution and decreases along with the increasing of $m=(\alpha,\beta)$, but the $N_3$ in the nonordered phase ($\lambda<1$) is robust to the spatial distribution for a properly chosen value of $\gamma=0.2$ (see the purple-solid lines in Fig. 4(c) and Fig. 4(d)). We further plot the $\tau^{\rm UB}_{\rm SEF}$ versus the parameter $\lambda$ for the spatial distributions $m=(3,3)$, $m=(4,4)$ in Fig. 4(e) and Fig. 4(f), which also illustrate that the anisotropic parameter ($\gamma=0.2$) can effectively modulate the range of the MQC distribution.

In order to elaborate on the modulation of spatial distribution of quantum correlations, we further analyze the correlation $N_3(\alpha,\beta)$ of reduced state $\rho_{ijk}$ in the nonordered and ordered phases, and compare the distribution of $N_3(\alpha,\beta)$ with that of concurrence $C(\alpha)$ in two-qubit state $\rho_{ij}$ with $\alpha$ being the distance between spins $i$ and $j$. In Fig. 5, we plot the spatial distributions of maxima of $N_3$ and $C$ along with the distance parameters $\alpha$ and $\beta$ in the two phases. As shown in Fig. 5(a), the maxima of $N_3(\alpha,\beta)$ versus the distribution indexes $(\alpha, \beta)$ in the nonordered phase ($\lambda<1$) is plotted, where the anisotropic parameter $\gamma$ is chosen to be $1.0, 0.6$, and $0.2$ in the three panels respectively. For the case of $\gamma=1$, the maximal distribution of $N_3$ is $m=(2,2)$ which coincides with the result for the Ising case given in Fig. 1(f). It should be noted that all the distributions are symmetric under swapping $\alpha$ and $\beta$ since the quantum state $\rho_{ijk}(\alpha,\beta)$ has the property. Along with the decreasing of the parameter $\gamma$, we find that the maximal distribution can attain to the longer ranges, where the $N_3$ in the second panel ($\gamma=0.6$) reaches to the distribution $m=(3,3)$ and the situation for $\gamma=0.2$ attains to $m=(7,7)$ and $(8,2)$ (we multiply the $N_3$ by a factor $2$ so as to show the trend of change more clearly in the third panel). In Fig. 5(b), we plot the  distribution of maxima of concurrence $C_{\rm max}(\alpha)$ along with the distance parameter $\alpha$ in the nonordered phase. For the Ising case ($\gamma=1$), the nonzero concurrence attains to the distribution $C_{\rm max}(2)$, and beyond the distance $\alpha=2$ the two-spin correlation disappears. Along with the decreasing of the parameter $\gamma$, the spatial distribution of $C_{\rm max}$ can reach to a longer range as shown in the second and the third panels, but the longest range $\alpha=6$ for $\gamma=0.2$ is still less than any one of the distance labels $(\alpha,\beta)$ for the nonzero $N_3(7,7)$ with $\gamma=0.2$, which means that the tripartite quantum correlation has the longer distribution range than that of two-qubit quantum correlation. In Figs. 5(c) and 5(d), we plot the distributions of $N_3$ and $C_{\rm max}$ in the ordered phase ($\lambda>1$), where the parameter $\gamma$ still can effectively modulate the distribution ranges of the two kinds of quantum correlations (we multiply the $N_3$ by a factor $1.5$ in the second panel of Fig. 5(c)). Moreover, in the ordered phase, the maxima of $N_3(\alpha,\beta)$ are always greater than those of $C_{\rm max}(\alpha)$ when the $\alpha$ is chosen to be the same value, for example, $N_3(12,1)\sim 10^{-3}$ but $C_{\rm max}(12)\sim 10^{-5}$ as shown in the third panels of Figs. 5(c) and 5(d). In addition, after comparing the distributions quantum correlations ($N_3$ and $C_{\rm max}$) in the two phases with $\gamma\neq 1$, we find that the spatial distributions in the ordered phase have the longer ranges than those in the nonordered case.

Our previous studies on the MQC distributions in the $XY$ model have been made in the case of thermodynamic limit. It is natural to ask whether or not the corresponding properties still hold for the case of finite chain length. We can investigate this problem by resorting to the fidelity of three-spin reduced states between the two cases \cite{nielsen00}
\begin{equation}\label{17}
	F(\rho_{ijk}^{(L)},\rho_{ijk}^{(\infty)})={\rm Tr} \left[ \sqrt{\sqrt{\rho_{ijk}^{(L)}}\rho_{ijk}^{(\infty)}\sqrt{\rho_{ijk}^{(L)}}}~\right],
\end{equation}
where $\rho_{ijk}^{(L)}$ is the reduced state for the finite chain length $L$, and $\rho_{ijk}^{(\infty)}$ is the one for $L\rightarrow \infty$. In Appendix C, we calculate the fidelities of the reduced states with $L=21$ for the spatial distributions $m=(1,1), (3,3), (6,3)$ and $(6,6)$ respectively, where the fidelities are larger than $0.99$ for most regions. Similar calculations are made for other distributions, and we can obtain the same qualitative results. The very high fidelities imply that the distribution properties obtained in the case of thermodynamic limit can still hold for the case of finite chain length ($L\geq 21$). The details for the analysis are presented in the Appendix.

\begin{figure}
	\epsfig{figure=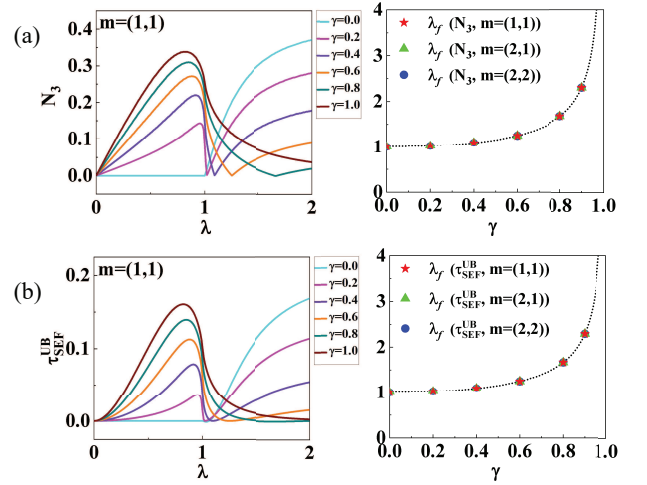,width=0.45\textwidth}
\caption{(Color online) The factorization property indicated by the sudden change of the tripartite quantum correlations $N_3$ and $\tau^{\rm UB}_{\rm SEF}$ in the thermodynamic limit. (a) Left panel: $N_3$ versus the parameter $\lambda$ for $\gamma=0,0.2,0.4,0.6,0.8,1.0$ with the spatial distribution $m=(1,1)$; Right panel: the comparison of the sudden change positions of $N_3$ (for $\gamma=0,0.2,0.4,0.6,0.8,0.9$) and the analytical result $\lambda_f=1/\sqrt{1-\gamma^2}$ (the black-dotted line) with the distributions $m=(1,1), (2,1)$ and $(2,2)$. (b) The similar plots for the correlation $\tau^{\rm UB}_{\rm SEF}$ with the left panel being the correlation evolution along with $\lambda$ and the right panel being the coincident transition positions with the analytical results.}
\end{figure}

Factorization is an important property of the $XY$ model, and we have shown in Figs. 4(a) and 4(b) that the tripartite quantum correlations $N_3$ and $\tau_{\rm SEF}^{\rm UB}$ are superior to the other two MQCs ($T_3$ and $\tau_{\rm SEF}^{\rm LB}$) because they can capture this interesting property. Here, we further analyze the factorization property in a more generic situation. For the case of three adjacent spins $m=(1,1)$ at the thermodynamic limit, the correlation $N_3$ versus the parameter $\lambda$ is plotted for different values of $\gamma$ in the left panel of Fig. 6(a), where the sudden change of the ground state can be indicated clearly by the zero $N_3$. In order to check the accuracy of the identification, we further plot these positions (the red stars for $m=(1,1)$) of sudden change and compare them with the analytical result $\lambda_f=1/\sqrt{1-\gamma^2}$ (the black-dotted line) in the right panel of this figure, which illustrates the nice identification on the factorization property. We also plot the sudden change positions for the spatial distributions $m=(2,1)$ (the green triangles) and $m=(2,2)$ (the blue circles) in this panel, which still exhibit a good coincidence. Similarly, the tripartite quantum correlation $\tau_{\rm SEF}^{\rm UB}(1,1)$ versus the parameter $\lambda$ is plotted in the left panel of Fig. 6(b), which can indicate the sudden changes of the ground state for different values of $\gamma$ too. Similarly, the accuracy of the identification on the factorization is illustrated by the right panel and the transition positions of $\tau_{\rm SEF}^{\rm UB}$ for the distributions $m=(1,1), (2,1)$ and $(2,2)$ are coincided with the analytical results.

\begin{figure}
	\epsfig{figure=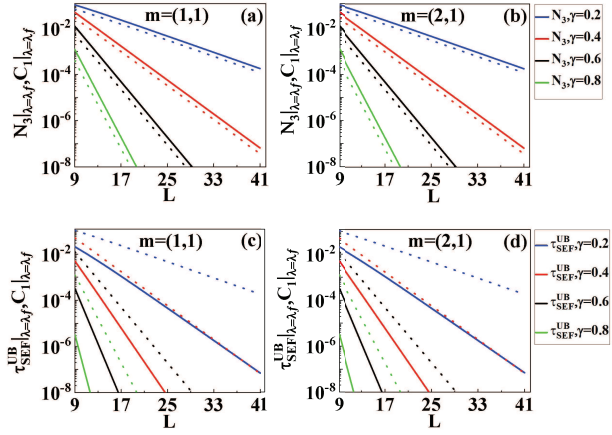,width=0.45\textwidth}
	\caption{(Color online) The factorization properties of tripartite quantum correlations $N_3$ and $\tau_{\rm SEF}^{\rm UB}$ for the finite chain length cases with the spatial distributions $m=(1,1)$ and $m=(2,1)$. (a) $N_3 (1,1)$ are plotted as the logarithmic functions of the chain length $L$ for the parameter $\gamma=0.2,0.4,0.6,0.8$ (the solid lines with different colors) and compared with the behaviors of the concurrence $C_1$ (the dotted lines with corresponding colors). (b) The similar plots of $N_3(2,1)$ at the factorization points. (c) and (d): The correlation $\tau_{\rm SEF}^{\rm UB}$ at the factorization points versus the chain length $L$ for different distributions.}
\end{figure}

It was pointed out that the position of the factorization point $\lambda_f=1/\sqrt{1-\gamma^2}$ is independent of the chain length in the $XY$ model \cite{mb10ps}, although the ground state at this point in the finite chain case may not be fully separable. At the factorization point with finite chain length, the ground state of the $XY$ system has a two-fold degeneracy, and the two eigenstates with even and odd parities can be written as \cite{smg13pra,smg08prl}
\begin{equation}\label{18}
	\begin{split}
		&\ket{\Phi_{\rm even}}=(\ket{\phi_+}+\ket{\phi_-})/N_+,\\
		&\ket{\Phi_{\rm odd}}=(\ket{\phi_+}-\ket{\phi_-})/N_-,
	\end{split}
\end{equation}
where $N_\pm=\sqrt{2(1\pm {\rm cos}^L \theta_i)}$ are the normalization coefficients and the two factorized components in each eigenstate have the form $\ket{\phi_\pm}=\otimes_{i=1}^{L}{\rm exp}(\frac{i}{2}\theta_\pm \sigma_{i}^{y})\ket{0}$ with $\theta_\pm=\pm {\rm arccos}(\sqrt{(1-\gamma)(1+\gamma)})$. Because the two factorized states $\ket{\phi_+}$ and $\ket{\phi_-}$ are nonorthogonal except for $\gamma=1$, the even and odd ground states given in Eq.~(18) are not orthogonal in general, which results in the ground state with $\gamma\neq 1$ being entangled at the factorization point. In Ref. \cite{smg13pra}, the authors studied the tripartite entanglement property at the factorization point in the case of finite chain length and found that the tripartite entanglement in the adjacent three spins obeys the finite-size scaling. In Fig. 7, we further investigate this factorization property of the tripartite quantum correlations $N_3$ and $\tau_{\rm SEF}^{\rm UB}$, which are plotted as the logarithmic functions of the chain length $L$ for the anisotropic parameter $\gamma=0.2,0.4,0.6,0.8$ (the solid lines with different colors in each panel) with the spatial distributions $m=(1,1)$ and $m=(2,1)$ and compared with the behaviors of the corresponding concurrence $C_1(\gamma)$ for the nearest neighbor two spins (the dotted lines with different colors in very panel). As shown in Fig. 7(a), we find the tripartite quantum correlation $N_3$ is nonzero at the factorization points and decreases exponentially with the increasing of chain length $L$, which obeys the similar finite-size scaling to that of the concurrence $C_1$ (the solid and dotted lines with the same color having the same value of $\gamma$). Moreover, $N_3$ is always greater than $C_1$ for the given values of $L$ and $\gamma$, and the value of $N_3$ decreases along with the increasing of $\gamma$ (for example, $N_3(\gamma=0.2)\sim 10^{-2}$ but $N_3(\gamma=0.6)\sim 10^{-6}$ when $L=21$). The similar property for $N_3$ with the spatial distribution $m=(2,1)$ is illustrated by Fig. 7(b), where we find the spatial distribution of three spins has little influence on the factorization property in comparison with the adjacent case $m=(1,1)$ in Fig. 7(a). In Figs. 7(c) and 7(d), we further plot the tripartite quantum correlation $\tau_{\rm SEF}^{\rm UB}$ with the distributions $m=(1,1)$ and $m=(2,1)$ in a similar way, where we find that the $\tau_{\rm SEF}^{\rm UB}$ still obeys the finite-size scaling and the spin distributions have little influence on the scalings but their decreasing exponents are different from that of two-qubit $C_1$(or $N_3$). For the given values of $L$ and $\gamma$, $\tau_{\rm SEF}^{\rm UB}$ is always less than $N_3$, and the different scaling exponents may come from the different measure methods for tripartite quantum correlation: $N_3$ quantifies the MQC by the way that the system is correlated in any bipartite partition, but $\tau_{\rm SEF}^{\rm UB}$ quantifies the MQC by the way that the correlation is multipartite which cannot be restored in any two-spin subsystems. It is noted that our analysis does not include the cases of $\gamma=0$ or $\gamma=1$ since the reduced states $\rho_{ijk}$ for the two cases are fully separable and then have zero tripartite quantum correlations \cite{smg13pra,el61ap}.

At the end of this section, we further analyze the difference between the two kinds of tripartite quantum correlations $N_3$ and $\tau_{\rm SEF}^{\rm UB}$ in the $XY$ model. For the given spatial distribution $m=(\alpha, \beta)$ and the values of parameters $\lambda$ and $\gamma$, the maximum of $N_3$ is greater than that of $\tau_{\rm SEF}^{\rm UB}$, which can be observed by the corresponding values in Fig. 4 and Fig. 6. In addition, for the case of finite chain length with the same parameters in Fig. 7, the correlation $N_3$ at the factorization point still has the larger value in comparison with that of $\tau_{\rm SEF}^{\rm UB}$. The larger value of $N_3$ makes its spatial distribution be well modulated via the parameter $\gamma$ as shown in Fig. 5. However, the detection ability of $\tau_{\rm SEF}^{\rm UB}$ for tripartite quantum correlation may be superior to that of $N_3$. In Ref. \cite{dp07njp}, the authors studied the bound entanglement \cite{phor97pla,mhor98prl} (the entangled state but with the positive partial transposition) of three spins in the $XY$ model at the thermodynamic limit, since the bound entangled state is a kind of useful physical resource for quantum secure communication \cite{khor05prl}. They found that, for the spatial distribution $m=(4,4)$ with the parameter $\gamma=0.5$, there is the bound entanglement in the reduced three-spin state $\rho_{i|jk}$ under the partition $i|jk$ when the parameter $\lambda$ is chosen in certain ranges. In Fig. 8, we plot the bipartite entanglement negativity $N(\rho_{i|jk})$ and tripartite quantum correlations $N_3$ and $\tau_{\rm SEF}^{\rm UB}$ as the functions of the parameter $\lambda$ for the given distribution and value of $\gamma$. We find that, when the parameter $0.959<\lambda<1.074$, both the bipartite negativity (the black-dotted line) and the tripartite quantum correlation $N_3$ (the red-solid line) are zero as shown in the figure, but the tripartite quantum correlation $\tau_{\rm SEF}^{\rm UB}$ (the blue-solid lines) has the nonzero value which indicates the existence of bound entanglement. According to the definition of $N_3$ in Eq. (3), this tripartite quantum correlation cannot identify the bound entanglement due to its construction coming from the product of bipartite negativities. Moreover, in the inset of Fig. 8, we further highlight the bound entanglement indicated by nonzero $\tau_{\rm SEF}^{\rm UB}$ in the parameter region $\lambda\in (1.205, 1.226)$.

\begin{figure}
	\epsfig{figure=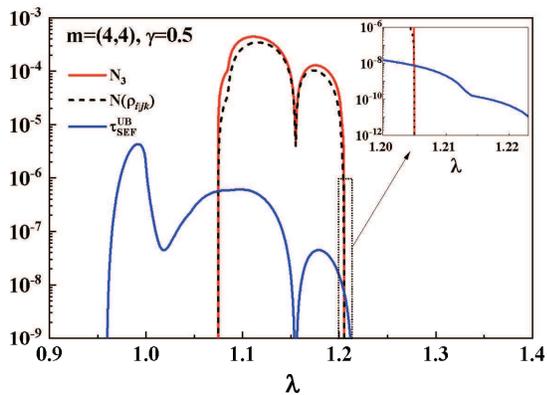,width=0.4\textwidth}
	\caption{(Color online) The bipartite entanglement negativity $N(\rho_{i|jk})$ (the black-dotted line) and tripartite quantum correlations $N_3$ (the red-solid line), $\tau_{\rm SEF}^{\rm UB}$ (the blue-solid line) as the functions of the parameter $\lambda$, where $\tau_{\rm SEF}^{\rm UB}$ has the stronger detection ability than that of $N_3$ and can indicate the bound entanglement. The inset highlight the bound entanglement detection when $\lambda\in (1.205, 1.226)$.}
\end{figure}

\section{Discussions and conclusion}

Based on multipartite correlation postulates and entanglement monogamy \cite{yco07pra,yk14prl,ben11pra}, the four utilized tripartite quantum correlations $N_3$, $T_3$, $\tau_{\rm SEF}^{\rm UB}$ and $\tau_{\rm SEF}^{\rm LB}$ we introduced in Sec. II are computable for an arbitrary three-qubit mixed state, which can be served as the basic tools for characterizing the properties of quantum many-body systems. In comparison with the tripartite quantum entanglement $I_2$ \cite{smg13pra} and genuine multipartite negativity \cite{mh14prb}, the tripartite quantum correlations have much larger spatial distributions, which makes our defined MQCs are more suitable to distinguish the quantum properties in the $XY$ system especially when the tripartite quantum entanglements and two-site concurrence are undetectable. In experiments, the observations on the distribution property of tripartite quantum correlations and the quantum critical phenomenon are possible in many-body systems. On the one hand, the observers only need process the data of several particles rather than complete information about the ground state; on the other hand, the experimental preparation about the quantum many-body system for dozens of qubits has been realized in superconducting platforms \cite{aru19nat,ye19prl,gong21sci}.

Hofmann \emph{et al} studied the four-partite entanglement in the $XY$ chain by using the genuine multipartite negativity, and found that the entanglement disappears when the distance between any two spins in the four-qubit subsystems is larger than $2$ \cite{mh14prb}. Moreover, by means of  the four-concurrence \cite{wong01pra}, Osterloh \emph{et al} analyzed the four-particle entanglement in the $XY$ chain, where the entanglement changes to zero when any two pairs of spins are next-nearest neighbor \cite{ao17pra}. It is meaningful to further investigate whether the four-partite quantum correlations have the larger spatial distributions. Based on the monogamy property of squared concurrence \cite{osb06prl}, a multipartite quantum correlation measure is presented in four-qubit systems \cite{bai07pra}, which can characterize the multipartite entanglement in cluster-class states \cite{baw08pra}. Particularly, the MQCs utilized in this paper can be easily extended to the four-qubit case and are computable for an arbitrary four-qubit mixed state. Besides the $XY$ model studied in this work, it is worthwhile to investigate the MQC modulation and critical property in other kinds of many-body systems such as Heisenberg spin chains with the alternating-field or Dzyaloshinskii-Moriya interaction \cite{tc16pra,tcy19prb}, multipartite quantum systems with topological quantum phases \cite{pez17prl}, multi-spin systems with long-range interactions \cite{ric14nat,sch21prr,jr22pre}, and so on.

In conclusion, we have studied the MQC properties and critical phenomena in a one-dimensional spin-1/2 $XY$ model, where the four tripartite quantum correlations we introduced are computable and can characterize effectively the MQC in the three-spin reduced state of the ground state. In the Ising case with $\gamma=1$, the spatial distribution of MQCs attains to the range $m=(3,2)$, which is larger than the maximal distribution of tripartite quantum entanglement $m=(2,1)$. All the correlations $N_3$, $T_3$, $\tau_{\rm SEF}^{\rm UB}$ and $\tau_{\rm SEF}^{\rm LB}$ can be used to detect the quantum phase transition and obey the finite-size scaling, even beyond the situation of nearest-neighbor three spins. Furthermore, we have shown that the two selected MQCs $N_3$ and $\tau_{\rm SEF}^{\rm UB}$ can capture exactly the sudden change behavior of the factorization property in the $XY$ system for both the thermodynamic and finite-size cases. In particular, it is revealed that the anisotropic parameter $\gamma$ can modulate effectively the spatial distribution of $N_3$ to a much longer range, which is very useful for the information propagation in multipartite systems. The similar modulation property still holds for the finite chain length according to the super high fidelity between the three-spin reduced states in the finite and infinite chains. In addition, the correlation $\tau_{\rm SEF}^{\rm UB}$ has more strong detection ability and can distinguish the bound entanglement in the $XY$ system when the tripartite quantum correlation $N_3$ loses its efficacy.

\section*{Acknowledgments}

This work was supported by  NSF-China (Grants Nos. 11575051 and 11904078), Hebei NSF (Grants Nos. A2021205020, A2019205266), the Key-Area Research and Development Program of GuangDong Province (Grant No. 2019B030330001), and project of China Postdoctoral Science Foundation (Grant No. 2020M670683). JR was also funded by Science and Technology Project of Hebei Education Department (Grant No. QN2019092).

\appendix
\section{The three-spin reduced density matrix of ground state in the $XY$ model}

In Sec. IIA, we introduced the Hamiltonian of the $XY$ model and gave the expression of three-qubit reduced density matrix $\rho_{ijk}$ of the ground state. Utilizing the translation invariance of the system, the reduced state of three spins with the spatial distribution $m=(\alpha,\beta)$ can be written as \cite{dp07njp}
\begin{equation}
\rho_{i-\alpha,i,i+\beta}=\frac{1}{8}{\underset{p,q,s}{\sum}}\langle\sigma^{p}_{i-\alpha}\sigma^{q}_{i}\sigma^{s}_{i+\beta}
\rangle_{\ket{\psi}}\sigma^{p}_{i-\alpha}\sigma^{q}_{i}\sigma^{s}_{i+\beta},
\end{equation}
where the label $i$ is an arbitrary spin position in the $XY$ chain, and the operators in the expectation value $\langle\sigma^{p}_{i-\alpha}\sigma^{q}_{i}\sigma^{s}_{i+\beta}\rangle_{\ket{\psi}}$ acts on the ground state with the summation indexes $p,q,s$ run over the set $\{x,y,z,0\}$.
According to this expression, the three-spin reduced density matrix can be obtained by calculating a set of correlations $\langle\sigma^{p}_{i-\alpha}\sigma^{q}_{i}\sigma^{s}_{i+\beta}\rangle_{\ket{\psi}}$. Combining the symmetry of the Hamiltonian, one can get some zero correlations, which results in half of the matrix elements of the three-qubit state being zero \cite{dp07njp}. After some derivations, we can obtain the reduced state in the computational basis
\begin{equation}
	\begin{split}
		&\rho_{i-\alpha,i,i+\beta}=\\
		&\frac{1}{8}\begin{pmatrix} a_{11} & 0 & 0 & a_{14} & 0 & a_{16} & a_{17} & 0 \\ 0 & a_{22} & a_{23} & 0 & a_{25} & 0 & 0 & a_{28} \\ 0 & a_{32} & a_{33} & 0 & a_{35} & 0 & 0 & a_{38} \\ a_{41} & 0 & 0 & a_{44} & 0 & a_{46} & a_{47} & 0 \\ 0 & a_{52} & a_{53} & 0 & a_{55} & 0 & 0 & a_{58} \\ a_{61} & 0 & 0 & a_{64} & 0 & a_{66} & a_{67} & 0 \\ a_{71} & 0 & 0 & a_{74} & 0 & a_{76} & a_{77} & 0 \\ 0 & a_{82} & a_{83} & 0 & a_{85} & 0 & 0 & a_{88} \end{pmatrix}.
	\end{split}
\end{equation}
The concrete form of the three-spin reduced density matrix will be available if we can calculate all the $32$ nonzero matrix elements, which is closely interrelated with the set of correlation functions $\{\langle\sigma^{p}_{i-\alpha}\sigma^{q}_{i}\sigma^{s}_{i+\beta}\rangle_{\ket{\psi}}\}$.
Using the method given in Refs.~\cite{el61ap,pp70ap,be71pra}, these three-spin correlations can be further decomposed into a series of determinants with elements only being two-spin correlation functions.
The aim of this Appendix is to give a brief description for the derivation and provide the detailed information needed to compute the three-spin reduced state in Eq. (A2).

Here, we first give a brief review on the derivation of characterizing three-spin correlation by the two-spin ones \cite{el61ap,pp70ap,be71pra}.
After Jordan-Wigner transformation, the Pauli operators in the Hamiltonian shown in Eq. (1) of the main text can be mapped into spinless fermion operators $c^{\dagger}_j$ and $c_j$, and we have the following relations
\begin{equation}
	\begin{split}
		&\sigma^x_l=A_l\underset{i=1}{\overset{l-1}{\prod}}A_iB_i,\\
		&\sigma^y_l=-iB_l\underset{i=1}{\overset{l-1}{\prod}}A_iB_i,\\
		&\sigma^z_l=-A_lB_l,
	\end{split}
\end{equation}
where the symbols denote $A_l=c_l+c^{\dagger}_l$ and $B_l=c^{\dagger}_l-c_l$ with the index $l$ being the  position of spin.
By using the Wick theorem, the three-site correlations can be decomposed into a series of products of three kinds of correlation functions $\langle A_lB_k\rangle_{\ket{\psi}}$, $\langle A_lA_k\rangle_{\ket{\psi}}$ and $\langle B_lB_k\rangle_{\ket{\psi}}$. After the exact diagonalization of the Hamiltonian, we can obtain \cite{el61ap}
\begin{equation}
	\begin{split}
		&\langle A_lA_k\rangle_{\ket{\psi}}=\delta_{lk}\\
		&\langle B_lB_k\rangle_{\ket{\psi}}=-\delta_{lk}\\
		&\langle A_lB_k\rangle_{\ket{\psi}}=G_{r},
	\end{split}
\end{equation}
where $\delta_{lk}$ is the delta function, and $G_r$ is a newly defined function with the subscript $r=k-l$ according to the property of translation symmetry. In the case of the thermodynamic limit, the new function has the form
\begin{equation}
	\begin{split}
		G_{r}=&\frac{1}{\pi}\int^{\pi}_0d\phi[\cos(\phi r)(1+\lambda\cos\phi)\\
		&-\gamma\lambda\sin\phi\sin(\phi r)]\frac{1}{\Lambda_{\phi}},
	\end{split}
\end{equation}
where the parameters are $\Lambda_{\phi}=\sqrt{\alpha^2+\beta^2}$, $\alpha=(\lambda\cos\phi+1)$, $\beta=\lambda\gamma\sin\phi$, respectively. In the case of finite chain with the length $L$, we have
\begin{equation}
	\begin{split}
		G_{r}=&\frac{1}{L}{\underset{q}{\sum}}\frac{1}{\Lambda_{q}}[\cos(\phi_q r)(1+\lambda\cos\phi_q)\\
		&-\gamma\lambda\sin\phi_q\sin(\phi_q r)],
	\end{split}
\end{equation}
where the parameter can be written as $\Lambda_q=\sqrt{\alpha^2_q+\beta^2_q}$, $\alpha_q=(\lambda\cos\phi_q+1)$, $\beta_q=\lambda\gamma\sin\phi_q$, and $\phi_q=2\pi q/L$, respectively.
Therefore, the nonzero matrix elements in Eq. (A2) can be calculated by the above method.

As an example, we give the detailed calculation procedure for the first matrix element $a_{11}$.
According to the expression in Eq.~(A1), the element $a_{11}$ in Eq. (A2) can be written as the sum of certain nonzero correlation functions
\begin{eqnarray}
		a_{11}&=&\langle III\rangle_{\ket{\psi}}+\langle \sigma^z_{i-\alpha}II\rangle_{\ket{\psi}}+\langle I\sigma^z_{i}I\rangle_{\ket{\psi}}\nonumber\\
		&&+\langle II\sigma^z_{i+\beta}\rangle_{\ket{\psi}}
		+\langle \sigma^z_{i-\alpha}\sigma^z_{i}I\rangle_{\ket{\psi}}+\langle \sigma^z_{i-\alpha}I\sigma^z_{i+\beta}\rangle_{\ket{\psi}}\nonumber\\
		&&+\langle I\sigma^z_{i}\sigma^z_{i+\beta}\rangle_{\ket{\psi}}+\langle\sigma^z_{i-\alpha}\sigma^z_{i}\sigma^z_{i+\beta}\rangle_{\ket{\psi}},
\end{eqnarray}
where the labels $i,\alpha, \beta$ in the subscript represent the spatial distribution of three spins, and $I$ is the identity operator of single qubit system.
For convenience, we further simplify the expression of Eq. (A7) by using some abbreviations for the single-, two- and three-spin correlations, for instance,
\begin{equation}
	\begin{split}
        &Z_{-\alpha}=\langle \sigma^z_{i-\alpha}II\rangle_{\ket{\psi}},\\
		&Z_{-\alpha}Z_{\beta}=\langle \sigma^z_{i-\alpha}I\sigma^z_{i+\beta}\rangle_{\ket{\psi}},\\
		&Z_{-\alpha}Z_0Z_{\beta}=\langle \sigma^z_{i-\alpha}\sigma^z_{i}\sigma^z_{i+\beta}\rangle_{\ket{\psi}},\\
	\end{split}
\end{equation}
where the subscripts $-\alpha$, $0$, and $\beta$ in the abbreviation means the relative distance to the spin $i$, and the letter $Z$ means the correlation is related to the Pauli operator of $z$ direction. In Eq.~(A7), it is clear that the element $a_{11}$ does not contains any correlation function related to Pauli operators $\sigma^x$ or $\sigma^y$. The reason is that both $\sigma^x$ and $\sigma^y$ are the anti-diagonal matrices, thus all the first elements of the three-spin correlation operator $\{\sigma^{p}_{i-\alpha}\sigma^{q}_{i}\sigma^{s}_{i+\beta}\}$ are zero when it contains $\sigma^x$ or $\sigma^y$.
According to the above method via Jordan-Wigner transformation and Wick theorem, we can calculate the matrix element $a_{11}$, and, after some derivation, we have
\begin{equation}
		\begin{split}
		a_{11}=&1-3G_0+3G^2_0-\!G_{\alpha}G_{-\alpha}\\
		&-G_{\alpha+\beta}G_{-\alpha-\beta}
		-G_{\beta}G_{-\beta}+Z_{-\alpha}Z_0Z_{\beta},\\
	\end{split}
\end{equation}
where the $G_r$ is the two-site correlation defined in Eq. (A4) with the subscript being the relative distance (for the case of infinite chain length, we use the expression in Eq. (A5), and we use the formula in Eq. (A6) when the chain has a finite chain length), and the three-spin correlation function can be decomposed into the determinant of a set of two-site correlations
\begin{equation}
	Z_{-\alpha}Z_0Z_{\beta}=-\begin{vmatrix} G_0 & G_{\alpha} & G_{\alpha+\beta} \\ G_{-\alpha} & G_0 & G_{\beta} \\ G_{-\alpha-\beta} & G_{-\beta} & G_{0} \end{vmatrix}.
\end{equation}

Similarly, we can calculate other nonzero matrix elements in Eq. (A2). It should be pointed out that, for all the three-spin correlations $\{\sigma^{p}_{i-\alpha}\sigma^{q}_{i}\sigma^{s}_{i+\beta}\}$ which concern $\sigma^x$ or $\sigma^y$, the effective contribution comes from the case that a pair of Pauli operators $\sigma^x$ or $\sigma^y$ appear in the three-spin correlation. After some derivations, we can obtain the formulas for other nonzero elements
\begin{equation}
\begin{split}
	 &\!a_{22}\!\!=\!\!1\!-\!G_0\!\!-\!G^2_0\!-\!G_{\alpha}G_{\!-\!\alpha}\!\!+\!G_{\alpha\!+\!\beta}G_{\!-\!\alpha\!-\!\beta}
	\!+\!G_{\beta}G_{\!-\!\beta}\!-\!\mathcal{A},\\
	 &\!a_{33}\!\!=\!\!1\!-\!G_0\!\!-\!\!G^2_0\!+\!G_{\alpha}G_{\!-\!\alpha}\!\!-\!G_{\alpha\!+\!\beta}G_{\!-\!\alpha\!-\!\beta}
	\!+\!G_{\beta}G_{\!-\!\beta}\!-\!\mathcal{A},\\
	 &\!a_{44}\!\!=\!\!1\!+\!G_0\!\!-\!G^2_0\!+\!G_{\alpha}G_{\!-\!\alpha}\!\!+\!G_{\alpha\!+\!\beta}G_{\!-\!\alpha\!-\!\beta}
	\!-\!G_{\beta}G_{\!-\!\beta}\!+\!\mathcal{A},\\
	&\!a_{55}\!\!=\!\!1\!-\! G_0\!\!-\!G^2_0\!+\!G_{\alpha}G_{\!-\!\alpha}\!\!+\!G_{\alpha\!+\!\beta}G_{\!-\!\alpha\!-\!\beta}
	\!-\!G_{\beta}G_{\!-\!\beta}\!-\!\mathcal{A},\\
	&\!a_{66}\!\!=\!\!1\!+\! G_0\!\!-\!G^2_0\!+\!G_{\alpha}G_{\!-\!\alpha}\!\!-\!G_{\alpha\!+\!\beta}G_{\!-\!\alpha\!-\!\beta}
	\!+\!G_{\beta}G_{\!-\!\beta}\!+\!\mathcal{A},\\
	&\!a_{77}\!\!=\!\!1\!+\! G_0\!\!-\! G^2_0\!-\!G_{\alpha}G_{\!-\!\alpha}\!\!+\!G_{\alpha\!+\!\beta}G_{\!-\!\alpha\!-\!\beta}
	\!+\!G_{\beta}G_{\!-\!\beta}\!+\!\mathcal{A},\\
	 &\!a_{88}\!\!=\!\!1\!+\!3G_0\!\!+\!3G^2_0\!\!-\!G_{\alpha}G_{\!-\!\alpha}\!\!-\!G_{\alpha\!+\!\beta}G_{\!-\!\alpha\!-\!\beta}
	\!-\!G_{\beta}G_{\!-\!\beta}\!-\!\mathcal{A},\\
	&a_{14}=X_0X_{\beta}+\mathcal{B}-Y_0Y_{\beta}-\mathcal{C},\\
	&a_{23}=X_0X_{\beta}+\mathcal{B}+Y_0Y_{\beta}+\mathcal{C},\\
	&a_{58}=X_0X_{\beta}-\mathcal{B}-Y_0Y_{\beta}+\mathcal{C},\\
	&a_{67}=X_0X_{\beta}-\mathcal{B}+Y_0Y_{\beta}-\mathcal{C},\\
	&a_{16}=X_{-\alpha}X_{\beta}+\mathcal{D}-Y_{-\alpha}Y_{\beta}-\mathcal{E},\\
	&a_{25}=X_{-\alpha}X_{\beta}+\mathcal{D}+Y_{-\alpha}Y_{\beta}+\mathcal{E},\\
	&a_{38}=X_{-\alpha}X_{\beta}-\mathcal{D}-Y_{-\alpha}Y_{\beta}+\mathcal{E},\\
	&a_{47}=X_{-\alpha}X_{\beta}-\mathcal{D}+Y_{-\alpha}Y_{\beta}-\mathcal{E},\\
	&a_{17}=X_{-\alpha}X_0+\mathcal{F}-Y_{-\alpha}Y_0-\mathcal{G},\\
	&a_{28}=X_{-\alpha}X_0-\mathcal{F}-Y_{-\alpha}Y_0+\mathcal{G},\\
	&a_{35}=X_{-\alpha}X_0+\mathcal{F}+Y_{-\alpha}Y_0+\mathcal{G},\\
	&a_{46}=X_{-\alpha}X_0-\mathcal{F}+Y_{-\alpha}Y_0-\mathcal{G},
\end{split}
\end{equation}
where the letters $X$ and $Y$ in the expressions imply that the correlation functions are from the expectation value of $\sigma_x$ and $\sigma_y$, and the other parameters have the forms $\mathcal{A}=Z_{-\alpha}Z_0Z_{\beta}$, $\mathcal{B}=Z_{-\alpha}X_0X_{\beta}$, $\mathcal{C}=Z_{-\alpha}Y_0Y_{\beta}$, $\mathcal{D}=X_{-\alpha}Z_0X_{\beta}$, $\mathcal{E}=Y_{-\alpha}Z_0Y_{\beta}$, $\mathcal{F}=X_{-\alpha}X_0Z_{\beta}$, and $\mathcal{G}=Y_{-\alpha}Y_0Z_{\beta}$, in which all the abbreviations can be expressed via two-site correlation functions
\begin{equation}
\begin{split}
	&Z_{-\alpha}Z_{\beta}=G_0^2-G_{\alpha+\beta}G_{-\alpha-\beta},\\
	&X_{-\alpha}X_{\beta}=(-1)^{\alpha+\beta}\begin{vmatrix} G_{-1} & \cdots & G_{\alpha+\beta-2} \\ \vdots &   & \vdots \\ G_{-\alpha-\beta} & \cdots & G_{-1} \end{vmatrix},\\
    &Y_{-\alpha}Y_{\beta}=(-1)^{\alpha+\beta}\begin{vmatrix} G_{1} & \cdots & G_{\alpha+\beta} \\ \vdots &   & \vdots \\ G_{-\alpha-\beta+2} & \cdots & G_{1} \end{vmatrix},\\
    &Z_{-\alpha}Z_0Z_{\beta}=-\begin{vmatrix} G_0 & G_{\alpha} & G_{\alpha+\beta} \\ G_{-\alpha} & G_0 & G_{\beta} \\ G_{-\alpha-\beta} & G_{-\beta} & G_{0} \end{vmatrix},\\
	&X_{-\alpha}X_0Z_{\beta}=\\
	&(-1)^{\alpha+1}\begin{vmatrix} G_{-1} & \cdots & G_{\alpha-2} & G_{\alpha+\beta-1} \\ \vdots &   & \vdots & \vdots \\ G_{-\alpha} & \cdots & G_{-1} & G_{\beta} \\ G_{-\alpha-\beta} & \cdots & G_{-\beta-1} & G_{0} \end{vmatrix},\\
    &Y_{-\alpha}Y_0Z_{\beta}=\\
	&(-1)^{\alpha+1}\begin{vmatrix} G_{1} & \cdots & G_{\alpha} & G_{\alpha+\beta} \\ \vdots &   & \vdots & \vdots \\ G_{-\alpha+2} & \cdots & G_{1} & G_{\beta+1} \\ G_{-\alpha-\beta+1} & \cdots & G_{-\beta} & G_{0} \end{vmatrix},\\
	&X_{-\alpha}Z_0X_{\beta}=\\
	&(-1)^{\alpha+\beta}\begin{vmatrix} G_{-1} & \cdots & G_{\alpha-2} & G_{\alpha} & \cdots & G_{\alpha+\beta-2} \\ \vdots &   & \vdots & \vdots &   & \vdots \\ G_{-\alpha+1} & \cdots & G_{0} & G_{2} & \cdots & G_{\beta} \\ G_{-\alpha-1} & \cdots & G_{-2} & G_{0} & \cdots & G_{\beta-2} \\ \vdots &   & \vdots & \vdots &   & \vdots \\ G_{-\alpha-\beta} & \cdots & G_{-\beta-1} & G_{-\beta+1} & \cdots & G_{-1} \end{vmatrix},\\
	&Y_{-\alpha}Z_0Y_{\beta}=\\
	&(-1)^{\alpha+\beta}\begin{vmatrix} G_{1} & \cdots & G_{\alpha-1} & G_{\alpha+1} & \cdots & G_{\alpha+\beta} \\ \vdots &   & \vdots & \vdots &   & \vdots \\ G_{-\alpha+2} & \cdots & G_{0} & G_{2} & \cdots & G_{\beta+1} \\ G_{-\alpha} & \cdots & G_{-2} & G_{0} & \cdots & G_{\beta-1} \\ \vdots &   & \vdots & \vdots &   & \vdots \\ G_{-\alpha-\beta+2} & \cdots & G_{-\beta} & G_{-\beta+2} & \cdots & G_{1} \end{vmatrix}.
\end{split}
\end{equation}

According to Eqs.~(A9)-(A12), one can obtain the concrete form of the three-spin reduced state $\rho_{i-\alpha,i,i+\beta}$, where the formulas of two-site correlation function $G_r$ in Eq. (A5) and Eq. (A6) are suitable for the finite and infinite chain cases, respectively.

\section{The critical phenomenon and finite-size effects in the Ising system indicated by other tripartite quantum correlations}

In Sec. III, utilizing the tripartite quantum correlation $\tau_{\rm SEF}^{\rm UB}$ with the spatial distribution $m=(1,1)$, we analyzed the critical phenomenon and finite-size effects in the Ising model. The MQC with the distributions $m=(2,1),(2,2)$ and $(3,1)$ can also be served as a good indicator to detect the QPT and characterize the finite-size effects. After some derivation, we can obtain the similar qualitative results and plots like Fig. 2 in the main text. Here, for the distribution $m=(3,1)$, we give the fitted relations for $\tau^{\rm UB (\infty)}_{\rm SEF}$ and $\partial_{\lambda}\tau^{\rm UB(L)}_{\rm SEF}$, which can be expressed as
\begin{eqnarray}
	&&\partial_{\lambda}\tau^{\rm UB (\infty)}_{\rm SEF}=0.0989 \ln|\lambda-\lambda_{\rm c}|+0.1240,\\
	&&\partial_{\lambda}\tau^{\rm UB(L)}_{\rm SEF}[\lambda_{\rm m}(L)]=-0.0989 \ln L+0.0814,
\end{eqnarray}
where the position of the minimum of $\partial_{\lambda}\tau^{\rm UB}_{\rm SEF}$ scales as $\lambda_{\rm m}(L)-\lambda_{\rm c}\sim L^{-1.20}$.

\begin{figure}
	\epsfig{figure=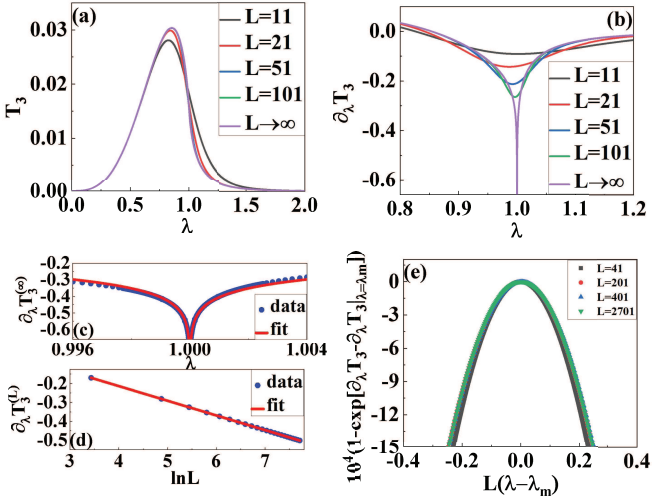,width=0.48\textwidth}
	\caption{(Color online) The critical phenomenon and finite-size scaling properties characterized by the tripartite quantum correlation $T_3(\rho_{ijk})$ with the spatial distribution $m=(1,1)$, including (a) the MQC with different chain lengths, (b) the derivative $\partial_{\lambda}T_3$ versus the parameter $\lambda$, (c) the fitting relation between $\partial_{\lambda}T_3^{(\infty)}$ and the parameter $\lambda$ close to the critical point, (d) the finite-size logarithmical scaling of $\partial_{\lambda}T_3^{(\rm L)}$, and (e) the homogeneous function for different chain lengths.}
\end{figure}

The tripartite quantum correlation $T_3$, we introduced in Sec. II of the main text, is also an effective indicator to characterize the critical phenomenon and finite-size effects in the spin model as shown in Fig. 9 where the spatial distribution of the MQC is chosen to be $m=(1,1)$. In Fig.~9(a), the three-spin correlation $T_3(\rho_{ijk})$ as a function of $\lambda$ is plotted with different chain lengths, and Fig.~9(b) shows the first-order derivative $\partial_{\lambda}T_3(\rho_{ijk})$ versus the parameter $\lambda$. Similar to the situation of $N_3$ and $\tau_{\rm SEF}^{\rm UB}$ shown in Fig.~2 and Fig.~3 of main text, we find that the curves for short chain lengths have a little deviation from the infinite chain case, and the $\partial_{\lambda}T_3(\rho_{ijk})$ diverges for infinite chain and have distinct minima for the finite chain lengths. Moreover, for the cases of infinite and finite chain length, the fitted functions of $T_3(\rho_{ijk})$ have the forms
\begin{eqnarray}
	&&\partial_{\lambda}T^{(\infty)}_3=0.0776 \ln|\lambda-\lambda_{\rm c}|+0.1304,\\
	&&\partial_{\lambda}T_3^{(L)}[\lambda_{\rm m}(L)]=-0.0776 \ln L+0.0968,
\end{eqnarray}
where $\lambda_{\rm m}(L)-\lambda_{\rm c}\sim -L^{-1.28}$ and the two fitted results are plotted in Fig.~9(c) and Fig.~9(d), respectively. Additionally, we can also give a general relation to exhibit the behavior of the tripartite quantum correlation $T_3$ in the quantum critical regime for finite system size,
\begin{equation}
	\partial_{\lambda}T_3-\partial_{\lambda}T_3|_{\lambda=\lambda_{\rm m}}=\mathcal{Q}_{T_3}[L(\lambda-\lambda_{\rm m})],
\end{equation}
where $\mathcal{Q}_{T_3}$ is a homogeneous function and plotted in Fig. 9(e) in which we utilize the data for chain sizes $L=41,201,401$ and $2701$, respectively.

\begin{figure}
	\epsfig{figure=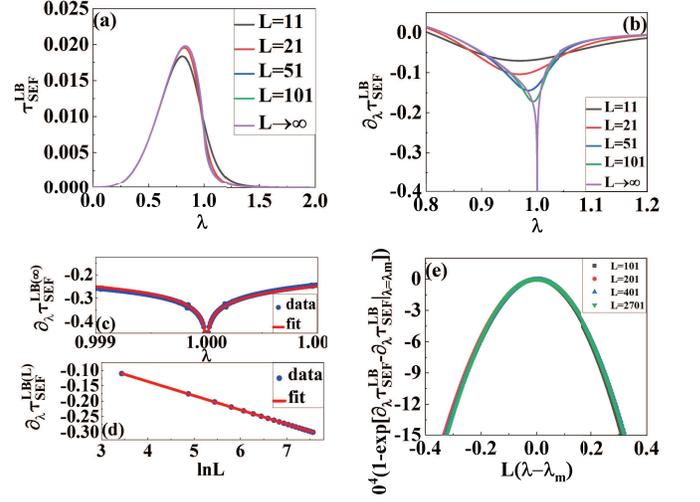,width=0.48\textwidth}
	\caption{(Color online) The critical phenomenon and finite-size scaling properties characterized by the tripartite quantum correlation $\tau^{\rm LB}_{\rm SEF}(\rho_{ijk})$ with the spatial distribution $m=(1,1)$, including (a) the MQC with different chain lengths, (b) the derivative $\partial_{\lambda}\tau^{\rm LB}_{\rm SEF}$ versus the parameter $\lambda$, (c) the fitting relation between $\partial_{\lambda}\tau^{\rm LB(\infty)}_{\rm SEF}$ and the parameter $\lambda$ close to the critical point, (d) the finite-size logarithmical scaling of $\partial_{\lambda}\tau^{\rm LB(L)}_{\rm SEF}$, and (e) the homogeneous function for different chain lengths.}
\end{figure}

The tripartite quantum correlation $\tau_{\rm SEF}^{\rm LB}$ with the distribution $m=(1,1)$ has the similar behaviors as shown in Fig. 10 and can characterize the critical phenomenon and finite-size effects. We plot the correlation versus the parameter $\lambda$ with different chain length $L$ in Fig. 10(a), and the derivation $\partial_{\lambda}\tau_{\rm SEF}^{\rm LB}$ can indicate the QPT as shown in Fig. 10(b). Similarly, $\tau^{\rm LB (\infty)}_{\rm SEF}$ and $\partial_{\lambda}\tau^{\rm LB(L)}_{\rm SEF}$ can be fitted with the relations
\begin{eqnarray}
	&&\partial_{\lambda}\tau^{\rm LB (\infty)}_{\rm SEF}=0.0461 \ln|\lambda-\lambda_{\rm c}|+0.0685,\\
	&&\partial_{\lambda}\tau^{\rm LB(L)}_{\rm SEF}[\lambda_{\rm m}(L)]=-0.0461 \ln L+0.0481,
\end{eqnarray}
which are plotted in Fig.~10(c) and (d). For the finite chain length, the minimum of $\partial_{\lambda}\tau^{\rm LB}_{\rm SEF}$ scales as $\lambda_{\rm m}(L)-\lambda_{\rm c}\sim L^{-1.32}$. Furthermore, the behavior of the tripartite quantum correlation $\tau^{\rm LB}_{\rm SEF}$ in the quantum critical regime for finite system size can be exhibited by a homogeneous function as shown in Fig. 10(e), where we utilize the data for chain sizes $L=41,201,401$ and $2701$, respectively.

\section{The fidelity of the three-spin reduced states between the finite and infinite $XY$ chains}

\begin{figure}
	\epsfig{figure=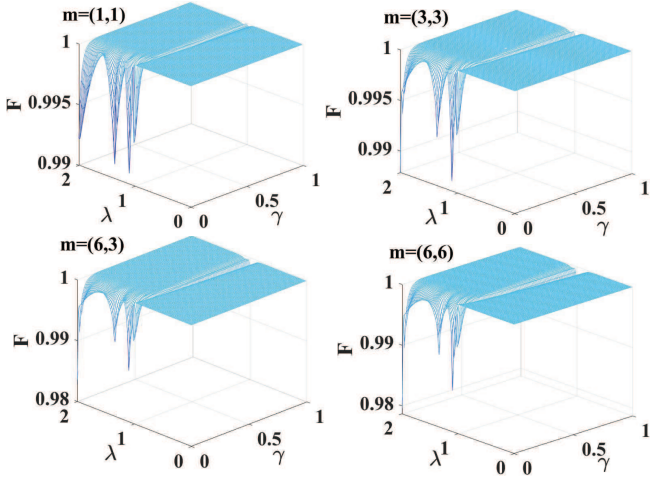,width=0.48\textwidth}
	\caption{(Color online) The fidelities of the reduced three-qubit states between the finite chain length $L=21$ and the infinite chain case. The four panels correspond to the spatial distribution $m=(1,1)$, $(3,3)$, $(6,3)$ and $(6,6)$ respectively.}
\end{figure}

In this Appendix, we will give the detail information of the calculated fidelities for the three-qubit reduced state between the finite chain length $L=21$ and infinite one with the spatial distributions of three spins being $m=(1,1), (3,3), (6,3)$ and $(6,6)$, respectively. According to the formula in Eq. (17) of the main text, we can calculate the fidelities with different distribution, which are plotted as the functions of parameters $\lambda$ and $\gamma$ in Fig. 11. As shown in the figure, it is obvious that the fidelities for most regions of  $\lambda$ and $\gamma$ are larger than 0.99, regardless of the spatial arrangement of three spins (we also calculated the fidelities for other spatial distributions and obtain the similar results).
That is to say, the difference between the three-qubit reduced density matrices in the finite and infinite chain cases is very small.
The quite high fidelities imply that the distribution properties of $N_3$ obtained in the thermodynamic limit ($L\rightarrow \infty$) still hold in the case of finite chain length. In particular, the anisotropic parameter $\gamma$ can modulate effectively the spatial distribution of $N_3$ to a longer range as shown in Fig. 5 of main text.

\end{document}